# Photophysics of Intrinsic Single-Photon Emitters in Silicon Nitride at Low Temperatures

*Zachariah O. Martin[1,2], Alexander Senichev[1,2], Samuel Peana[1,2], Benjamin J. Lawrie[2,3], Alexei S. Lagutchev[1,2], Alexandra Boltasseva[1,2], and Vladimir M. Shalaev[1,2*]*

[1]*Elmore Family School of Electrical and Computer Engineering, Birck Nanotechnology Center and Purdue Quantum Science and Engineering Institute, Purdue University, West Lafayette, IN, 47907, USA*

[2]*Quantum Science Center, Department of Energy, A National Quantum Information Science Research Center of the U.S., Oak Ridge National Laboratory, Oak Ridge, TN 37830, USA*

*Materials Science and Technology Division, Oak Ridge National Laboratory, Oak Ridge, TN 37830, USA*

Author email address: shalaev@purdue.edu

**Abstract:** A robust process for fabricating intrinsic single-photon emitters in silicon nitride has been recently established. These emitters show promise for quantum applications due to room-temperature operation and monolithic integration with the technologically mature silicon nitride photonics platform. Here, the fundamental photophysical properties of these emitters are probed through measurements of optical transition wavelengths, linewidths, and photon antibunching as a function of temperature from 4.2K to 300K. Important insight into the potential for lifetime-limited linewidths is provided through measurements of inhomogeneous and temperature-dependent homogeneous broadening of the zero-phonon lines. At 4.2K, spectral diffusion was found to be the main broadening mechanism, while time-resolved spectroscopy measurements revealed homogeneously broadened zero-phonon lines with instrument-limited linewidths.

## 1. Introduction

Single-photon sources form a critical building block for quantum information technologies such as all-optical quantum computing schemes, quantum communication protocols, quantum sensing, and quantum memories.[1–7] Heralded single photon sources utilizing nonlinear processes such as spontaneous parametric down conversion (SPDC) or four-wave mixing can produce single-photons with near-unity indistinguishability and have been used for ground-breaking demonstrations of satellite-based quantum communication[8,9] and quantum computational advantage.[10,11] One of the drawbacks of the demonstrated heralded nonlinear sources is the probabilistic nature of single-photon generation, which results in substantial obstacles to scalable quantum technologies. [12,13] In contrast, single photon emission from localized excitons or color centers in solids can be deterministic with sufficient control of the excitation and collection efficiency.[7,14] Much research attention has been given to the search for solid-state single photon emitters (SPEs) that can emit indistinguishable photons on-demand into desired optical modes. The nitrogen vacancy (NV) and silicon vacancy (SiV) centers in diamond, single quantum dots (QDs), and the various color centers in hexagonal boron nitride (hBN) have been rapidly developed in recent years for this purpose.[1,3,5–7,15–18] Moreover, these solid-state SPE's promise integrability with large scale on-chip quantum photonic circuitry.[2,4,15,18–41]

Researchers have made substantial advances in hybrid integration approaches, where an emitter in one material couples to a photonic element made of another material.[24,25,28,30,32,40,41] Since some materials that host SPEs cannot readily serve as photonic elements – due to intrinsic losses, difficulties in growing wafer-scale films, or nanofabrication challenges – these approaches enabled the on-chip integration of solid-state SPEs with favorable optical properties.[22,26–28,30,32,33,38,39,41–49] Drawbacks of hybrid integration, however, include increased device fabrication complexity and coupling losses at the interface between the SPE host material and the photonic element. Consequently, growing research efforts have targeted new species of color centers in materials that can also serve as a photonic platform.[1–4,34–36,47–54] This monolithic integration approach avoids the fabrication complexity and coupling losses of hybrid integration schemes. Promising results with new defect species in photonic materials include that of the carbon-based G-center and other emitters in silicon, color centers in gallium nitride, color centers in aluminum nitride, and the various defect centers in silicon carbide.[47–49,55–66] In addition to having outstanding optical properties, these defects have been monolithically integrated into photonic structures such as waveguides or 1D photonic crystals with high coupling efficiency.[24,48,49,52–54,60,67–69]

We have recently discovered a new type of SPE in SiN films, which exhibit high single-photon purity at room temperature and promise seamless integration with a technologically mature material platform.[70] These intrinsic SiN quantum emitters are stable, bright, and linearly polarized, which makes them exciting candidates for quantum photonic applications.[70] Furthermore, this development is of particular technological interest due to the low losses in SiN and mature fabrication techniques established for SiN photonic circuit elements.[71,72] SiN has become a leading material platform for quantum photonic computing and is employed by industry leaders and emerging photonic companies such as Xanadu and QuiX Quantum.[11,73,74] We have also demonstrated monolithic integration of SiN color centers into waveguides, as well as deterministic placement of emitters by patterning nanopillar arrays to explore the prospects of native SPEs in this platform. [50,75]

At room temperature SiN SPEs produce a broad photoluminescence (PL) spectrum that consists of multiple emission peaks. We found that for most SPEs these peak positions clustered around specific wavelengths, which suggests that single-photon emission originated from a particular type of defect center. However, the PL spectra reported lacked a distinct zero phonon line (ZPL) peak, and the nature of the optical transitions responsible for the broad spectral components was unclear.[70] Moreover, as we have previously shown, understanding and controlling additional spectral lines in SPEs is critical to their ultimate use in quantum technologies.[76,77] Here, we study the optical transition wavelengths, linewidths, and photon antibunching properties of these emitters as a function of temperature from 4K to 300K. At cryogenic temperatures, we find the emergence of narrow emission lines on top of the broad spectrum, which allowed us to identify ZPL peaks. The linewidths of these peaks are about 0.3 nm at 4.2K and exhibit inhomogeneous broadening as evidenced by their Gaussian shape. Time-resolved spectral measurements allow us to reveal the spectral diffusion responsible for inhomogeneous broadening and isolate homogeneously broadened peaks whose linewidths of ~0.03 nm are limited by the spectrometer resolution.

## 2. Results and Discussion

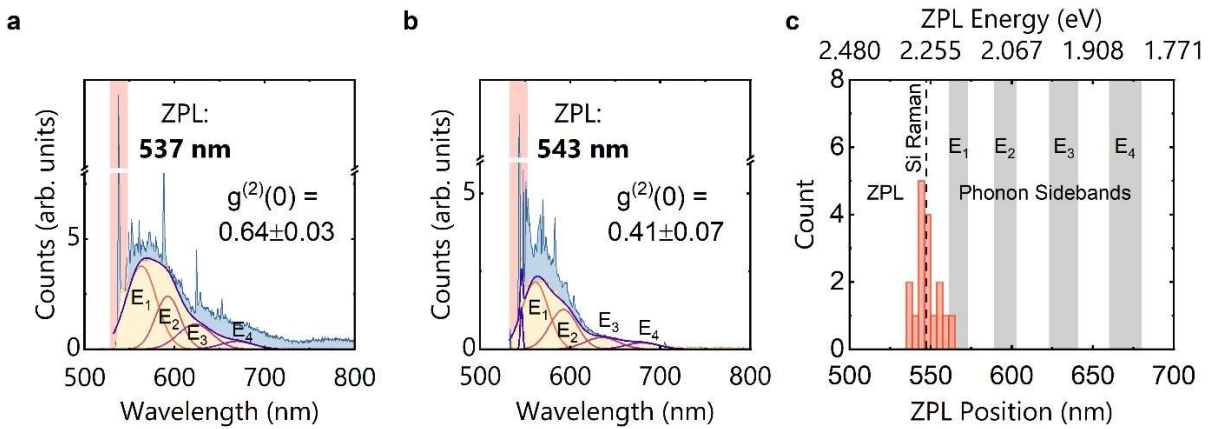

**Figure 1:** a,b) Background-corrected PL spectra taken from representative emitters at 4.2K (blue) and room temperature (yellow). The 4.2K and room-temperature PL intensities are scaled for comparison. c) Histogram of ZPL peak position acquired from 17 emitters. Shaded regions labeled $E_1$, $E_2$, $E_3$, and $E_4$ indicate the position of broad peaks typically resolved in room temperature spectra. [70]

Following the emitter creation procedure established previously[70], we fabricated SiN films that contain intrinsic single photon emitters. We first compare 4.2K and room-temperature PL spectra from SPEs in these SiN films which have been background corrected to emphasize the contribution from the emitters themselves (Figure 1a and 1b). The PL spectra observed at room temperature from isolated emitters agree well with our previous report[70] and are comprised of broad peaks spanning 550 nm – 800 nm (Figure 1a and 1b, yellow). However, as the sample is cooled below 200K, sharp lines appear on top of these broad peaks (Figure 1a and 1b, blue). We tentatively assign the broad peaks centered around 567 nm, 596 nm, 632 nm, and 670 nm to phonon sidebands (PSBs) and the lowest wavelength narrow peak to the emitter's (ZPL). The origin of the redshifted sharp peaks that sit atop the broad spectral features will be discussed below. From low-temperature spectra, we find an average ZPL position of 548 nm, with a standard deviation of around 7 nm (Figure 1c). Thus, we conclude that the lack of a clear ZPL transition in the previously reported SPE spectra[70] is a result of a large Huang-Rhys factor that leads to strong PSBs relative to the ZPL at room temperature (a 550 nm longpass filter that was used to acquire the previously reported spectra also minimized the appearance of the ZPL).

We also note that ZPL wavelengths from SPEs in SiN tend to cluster around the Si Raman line at 547.2 nm (520 cm$^{-1}$) under 532 nm excitation. The ZPL emission is clearly distinct from the Si Raman line though, as it only appears at the positions of localized emitters whereas the Si Raman peak is a feature of the underlying Si substrate and present everywhere (see Figure S1 and S2). Further, the ZPL emission wavelength and linewidth show clear temperature dependence, as will be discussed below, but the Si Raman line position and linewidth does not (see Figure S3).

Second-order autocorrelation ($g^{(2)}(\tau)$) measurements confirmed that the emitters analyzed here are single-photon sources. Note that we observed somewhat larger $g^{(2)}(0)$ values here than in our previous report[70], which we attribute to a larger beam spot size used for low temperature PL measurements. Statistics on the $g^{(2)}(0)$ distribution obtained are presented in Figure S1.

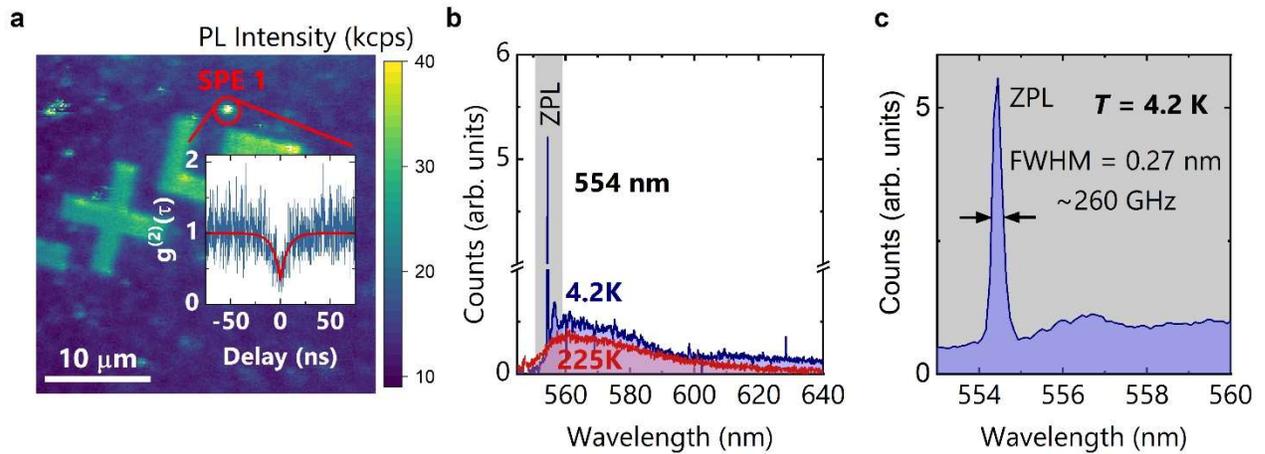

**Figure 2:** Characterization of a representative SPE in SiN at cryogenic temperatures. a) PL intensity map of a SiN sample containing SPEs. The red circle indicates the position of the representative emitter characterized in (b,c). The inset shows the emitter's second-order autocorrelation histogram, with an antibunching value of $g^{(2)}(0) = 0.41 \pm 0.04$. b) Background-corrected room temperature (red) and T=4.2K (blue) spectra of the representative SPE with the zero-phonon line (ZPL) highlighted in the low temperature spectrum. c) Magnified view of the ZPL region in (b).

Quantifying the linewidths of the peaks discussed earlier is of critical importance for defining the rate of dephasing processes and assessing the possibility of indistinguishable photon generation from SiN SPEs. PL intensity maps reveal the presence of isolated bright spots (Figure 2a), which were confirmed to be single-photon emitters by second-order autocorrelation histogram measurements (inset of Figure 2a). A representative emitter is indicated by the red circle in Figure 2a. Comparison of the emitter's room temperature and low temperature spectra (shown in red and blue, respectively, in Figure 2b) again reveals the emergence of a sharp ZPL at 4.2 K. This ZPL can be well fitted by a Gaussian peak with a linewidth of 0.27 nm or ~260 GHz (Figure 2c), which is spectrometer limited (the spectrometer resolution was selected to provide access to the ZPL and PSBs in a single spectrum). Greater spectral resolution is therefore needed to better ascertain the ZPL linewidth at cryogenic temperatures. The Gaussian shape of this line at 4.2 K also suggests that the emitter is affected by some inhomogeneous broadening mechanism such as spectral diffusion.

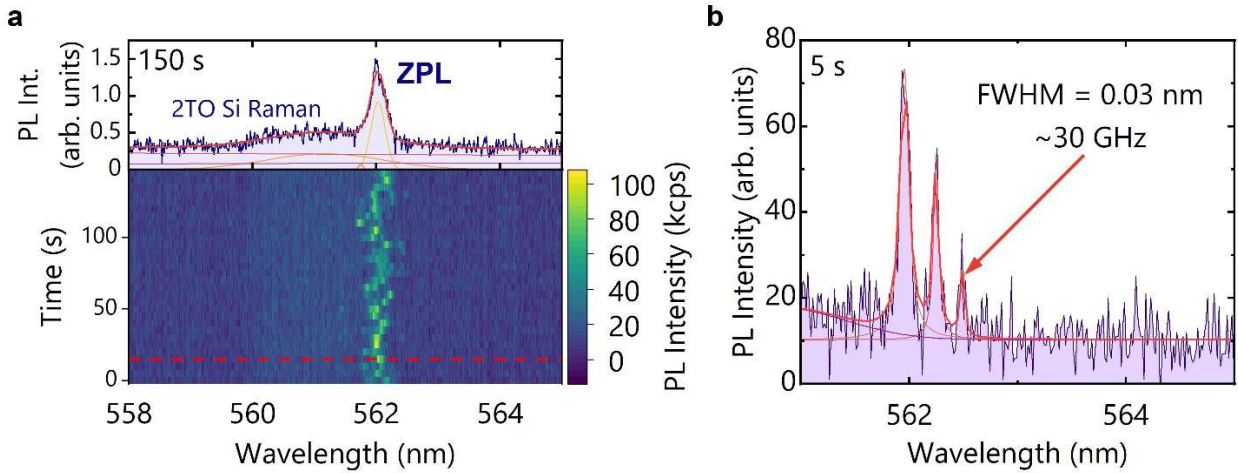

**Figure 3:** a) Time-resolved spectra from another emitter taken at 4.2 K using 0.029 nm resolution. The top plot shows the time-integrated (integrated for 150 s) spectrum. b) An individual frame of the time-resolved (integrated for 5 s) spectrum indicated by the dashed red line in (a).

To fully assess inhomogeneous linewidth broadening, we measured time-resolved spectra at higher spectral resolution. The time-resolved spectrum for an emitter with a ZPL around 562 nm shows substantial spectral diffusion, as the ZPL peak position wanders erratically over the course of the measurement (bottom of Figure 3a). The $2^{nd}$-order transverse optical (2TO) Si Raman mode appears in the time-resolved spectrum since it is not background-corrected (labeled in the top of Figure 3a). Fitting the time-integrated spectrum results in a linewidth of 0.28 nm, which is not limited by the spectrometer resolution (0.029 nm). However, analysis of a single frame that was integrated for 5 s (Figure 3b) reveals the presence of multiple, narrower ZPL peaks. These narrow peaks fit well to Lorentzian functions and have linewidths of $0.10 \pm 0.01$ nm, $0.06 \pm 0.01$ nm, and $0.03 \pm 0.01$ nm (~90 GHz, ~55 GHz, and ~30 GHz). The narrowest of these peaks is close to the spectrometer resolution limit; additional frames from time-resolved spectra also reveal instrument-limited linewidths, shown in Figure S5. Quantifying the indistinguishability of these emitters beyond this limit will require photoluminescence excitation spectroscopy or Hong-Ou-Mandel interferometry measurements, for instance. We also find that the rate and range of spectral wandering decreases at lower excitation power, which could potentially be used to control spectral diffusion (see Supplementary Section 4).

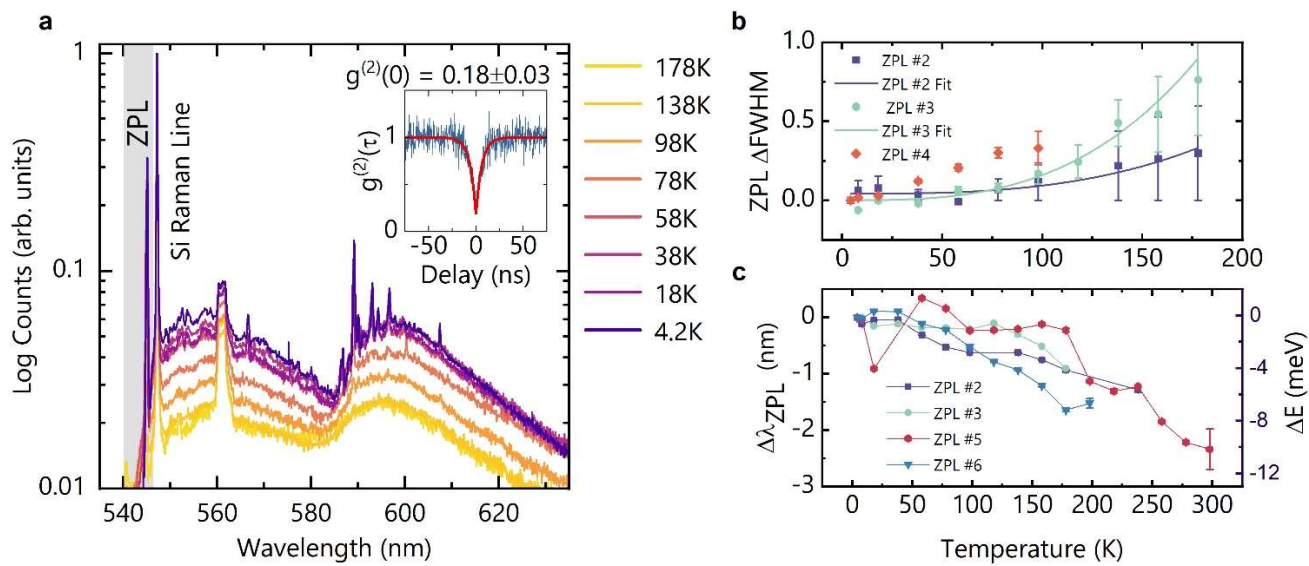

**Figure 4:** Temperature-dependent PL spectra analysis. a) Gradual emergence of narrow linewidth peaks in the SPE's PL spectrum with decreasing temperature (log scale, without background correction). Inset: second-order autocorrelation histogram recorded at 4.2 K confirming the single-photon nature of the emitter. b) Increase in the ZPL linewidth with increasing temperature from fitting temperature-dependent spectra of additional emitters. c) Change in ZPL energy of selected emitters from 4.2 K to 300 K.

Finally, we measure PL spectra and extract peak positions and linewidths over a range of temperatures from 4.2 K to 300 K. From taking temperature-dependent PL spectra from an additional representative emitter, we see that as the temperature decreases, the broad emission peaks seen at room temperature are preserved, and sharp emission lines gradually begin to emerge on top of them (Figure 4a; note that spectra are not background corrected and a logarithmic scale is used). Additionally, Figure 4b and 4c reveal the temperature dependent ZPL broadening and energy shift from fitting spectra taken from several different emitters (photophysical properties of selected representative emitters are tabled in Table S1). In some cases, the increase in ZPL linewidth can be well fitted with a $T^3$ dependence, which is similar to the behavior of other solid-state emitters reported in the literature.[65,78] The linewidth reaches 1.45 THz at ~200 K, and the ZPL peak position exhibits an overall redshift with increasing temperature. The ZPL energies of some emitters also behave quite nonmonotonically as a function of temperature from ~ 4 K to ~ 150 K (Figure 4c). However, the nature of the trend shown in Figure 4c is not clearly reproducible between emitters and will require further investigation in follow-up studies.

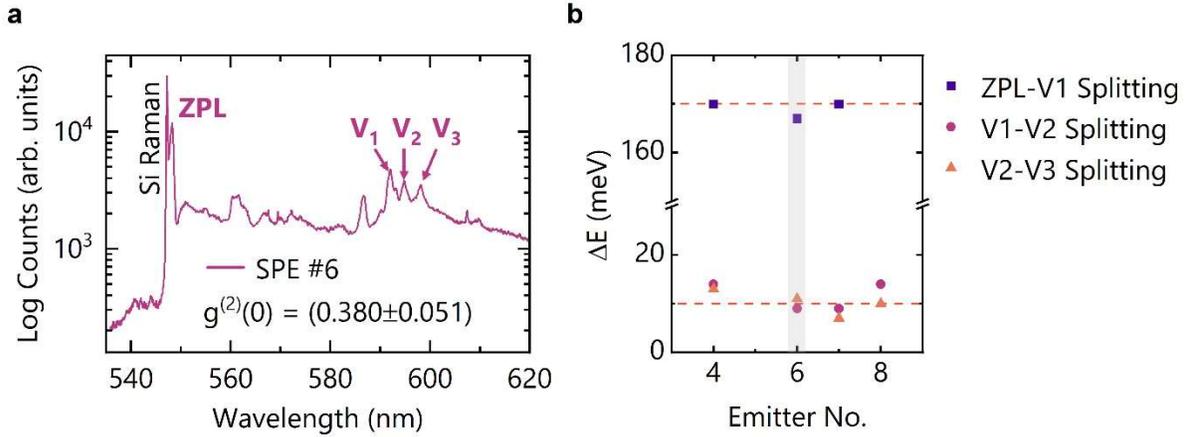

**Figure 5:** a) 4.2K spectra of a representative emitter highlighting the presence of a ZPL and redshifted sharp peaks. b) Comparison of the energy splitting between the ZPL and vibronic peaks for several representative ZPLs, as well as the energy differences between the vibronic peaks. The emitter whose spectra is shown in part (a) is highlighted.

As discussed previously, the low temperature PL spectra of SiN SPEs consist of a ZPL, PSBs, as well as additional previously unobserved narrow emission lines. We present low temperature spectra from an additional representative emitter (Figure 5a) to highlight this finding. The autocorrelation value at zero time-delay $g^{(2)}(0) = 0.38 \pm 0.05$ taken from the same location indicates this spectrum corresponds to that of a single emitter. In addition to the ZPL, we label the three additional peaks near 600nm $V_1$, $V_2$, and $V_3$. $V_1$ is redshifted from the ZPL by 167 meV, while $V_1$ and $V_2$ and $V_2$ and $V_3$ are split by 9 meV and 11 meV, respectively. From performing the same analysis on additional representative emitters with $g^{(2)}(0) < 0.5$ we find that the ZPL-$V_1$ splitting averages to around 170 meV, while the splitting between all vibronic peaks is on the order of 10 meV (Figure 5b). A similar structure was observed in low-temperature spectra of both single organic molecules[79] and optically active defects formed at the interface of silicon carbide (SiC) and $SiO_2$.[80] In the case of single organic molecules, sharp lines that are red-shifted from the ZPL also sit atop a broad peak; these lines correspond to localized vibronic modes of the crystal that the electronic transition couples to. The splitting between vibronic peaks of 5-10 meV observed in single dibenzoterrylene (DBT) molecules is also similar to what we observe here.[79] At 4.2 K, the $SiC/SiO_2$-related defects exhibit a pronounced ZPL with a group of additional red-shifted narrow lines, which were assigned to various components of the vibronic sidebands. Due to the defects' proximity to the interface, it was argued that these sidebands consist of both SiC and $SiO_2$ intrinsic phonons.[80] The splitting between the ZPL and the vibronic sidebands was also found to be between 150 and 220 meV, which similar to the splitting observed in our work on the $SiN/SiO_2$ system.[80] Therefore, we hypothesize that the additional sharp peaks observed in low temperature SiN SPE spectra could originate from vibronic modes.

The findings of our work further corroborate that the observed SPEs in SiN are related to a single defect family. The rich structure of the vibronic modes indicate that the defect structure could be more complex than defects existing within Si-Si or N-N bonds and may also include oxygen from the $SiO_2$ interface. It is also possible that partial crystallization of the SiN layer at the $SiO_2$ interface which occurs during annealing contributes to the presence of these vibronic modes.[81–83] Though low-temperature PL measurements have revealed rich spectra, follow up materials characterization studies supported by PL spectroscopy and theoretical modeling are needed to better understand the physical properties of the SiN films and shed light on the SPE defect structure.

## 3. Conclusion

In summary, we study the temperature evolution of the photophysical properties of SiN SPEs. We find sharp emission lines appear on top of broad peaks in SiN SPE PL spectra as emitters are cooled from room to cryogenic temperatures. At 4.2K, the ZPL emission occurs around 548 nm, and its linewidth is predominantly determined by spectral diffusion. We investigate this spectral diffusion in time-resolved PL spectroscopy and find narrow peaks whose linewidths are limited by the spectrometer resolution. Though inhomogeneous broadening remains a hurdle, our results show that there is still potential to obtain lifetime-limited emission peaks at low temperature for quantum photonic applications. This would open the possibility of quantum photonic circuitry with elements containing monolithically integrated SPEs. Future work on SiN SPEs will focus on mitigating spectral diffusion at cryogenic temperatures to the level required for the generation of indistinguishable photons. At non-cryogenic temperatures, we will explore using plasmonic and photonic cavities to enhance the spontaneous emission rate beyond the dephasing rate. Finally, we will work on a thorough materials investigation of the $SiN/SiO_2$ interface to better understand the defect formation mechanisms and atomic structure.

## 4. Experimental

### 4.1 PL setup

Room temperature optical characterization was performed using a custom-made scanning confocal microscope with a 100 μm pinhole based on a commercial inverted microscope body (Nikon Ti−U). To perform PL intensity mapping, a P-561 piezo stage driven by an E-712 controller (Physik Instrumente) scanned a 100x air objective with a 0.95 numerical aperture (NA). A continuous wave 532 nm laser (RGB Photonics) was used for excitation. The excitation beam was reflected off a 550 nm long-pass dichroic mirror (DMLP550L, Thorlabs), and a 550 nm long-pass filter (FEL0550, Thorlabs) was used to filter out the remaining pump power. Two avalanche detectors with a 30 ps time resolution and 35% quantum efficiency at 650 nm (PDM, Micro-Photon Devices) were used for single-photon autocorrelation measurements. Time-correlated photon counting was performed by a correlation card "start-stop" acquisition card with a 4 ps internal jitter (SPC-150, Becker & Hickl).

Low temperature PL characterization of emitters was performed using a home-built confocal PL microscope integrated into a closed-cycle Montana S100 cryostation. A Princeton Instruments Isoplane SCT-320 spectrograph with a Pixis 400BR Excelon camera was used to measure PL spectra. A grating with 2400 grooves/mm was used to measure ZPLs in Figure 3 with spectral resolution of 0.029 nm, and a grating with 600 grooves/mm was used to measure the broader spectra in Figure 2, 4, and 5 with spectral resolution of 0.13 nm. For direct comparison of room temperature and low temperature spectra over most of the visible range in Figure 1, a grating with 150 grooves/mm and spectral resolution of 0.55 nm was used. A 532 nm Cobolt diode laser was used to excite the SPEs via a 100x in-vacuum objective (Zeiss, NA = 0.85). A 2-axis galvo scanner integrated with a 4f imaging system was used to acquire PL spectrum images. Photon antibunching measurements were acquired concurrently with PL spectra via a 90:10 non-polarizing beam splitter that delivered 90% of the PL to a 50:50 fiber beamsplitter and a pair of large-area superconducting nanowire single-photon detectors. A Picoquant Hydraharp time-correlated single photon counting (TCSPC) system was used to timetag detected photon events and calculate the histogram of photon coincidences.

### 4.2 Growth and annealing

Samples containing SiN single photon emitters were prepared by a similar process to that reported previously.[50,70] First, $SiO_2$ films were grown on top of commercially available Si substrates by HDPCVD (Plasma-Therm Apex SLR). The SiN layer was then grown on top of the oxide layer by the same process. As in previous reports, we employed a nitrogen-rich growth recipe to form the SiN layer. The ratio of precursors $N_2$ to $SiH_4$ was set to 1.74, producing a non-stochiometric film with low auto-fluorescence in the visible region. The samples then underwent rapid thermal annealing (RTA) at 1100C for 120s in a Jipelec Jetfirst RTA system immediately after growth to activate SPE's. This procedure could reliably produce bright, stable, linearly polarized and high-purity SPEs in SiN films with low background autofluorescence.

**Supporting Information**

**Acknowledgements**

The cryo-optical spectroscopies described here were supported by the U.S. Department of Energy, Office of Science, National Quantum Information Science Research Centers, Quantum Science Center. The sample preparation, material and optical characterization were supported by National Science Foundation (NSF) grant 2015025-ECCS and Purdue's Elmore ECE Emerging Frontiers Center "The Crossroads of Quantum and AI." Low temperature photoluminescence measurements were conducted as part of a user project at the Center for Nanophase Materials Sciences (CNMS), which is a US Department of Energy, Office of Science User Facility at Oak Ridge National Laboratory.

**Data Availability**

All data is available upon reasonable request.

**Conflict of Interest**

The authors have no conflicts of interest to report.

# Supporting Information

## 1. Photon Antibunching Value Statistics

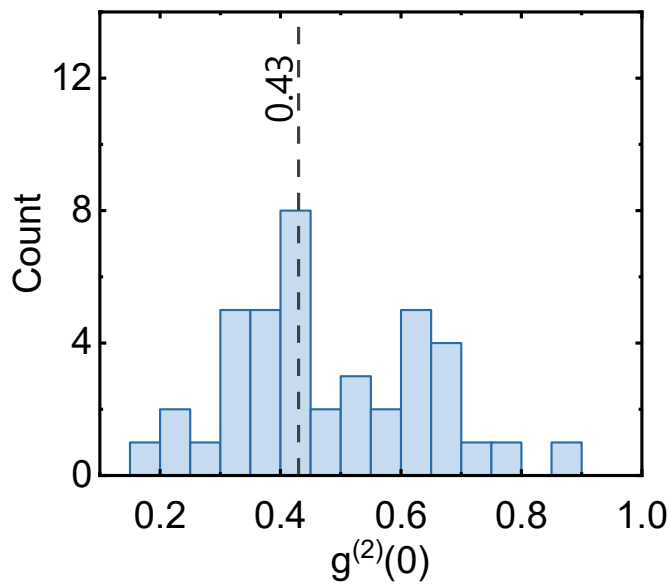

**Figure S1:** Distribution of $g^{(2)}(0)$ values taken at low temperature from 41 emitters. The median $g^{(2)}(0)$ value of 0.43 is indicated by the dashed line. The bin size is 0.05

The distribution of second-order correlation histogram values at zero time delay $g^{(2)}(0)$ taken at cryogenic temperatures from 41 emitters is plotted in Figure S1. We find a median $g^{(2)}(0)$ value of 0.43, which is indicated by the dashed line. The single-photon purity found here is lower than in our first report.[70] We attribute this difference to the larger excitation beam spot size used in the cryogenic measurement setup inducing increased background fluorescence. However, we note that the median $g^{(2)}(0)$ value is still well below 0.5.

## 2. Background and Emitter Spectra Comparison

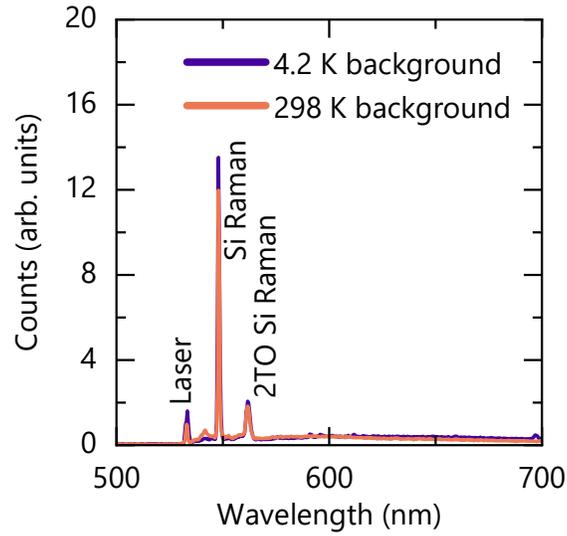

**Figure S2:** Comparison of 4.2K (purple) and 238K (yellow) background spectra.

Figure S2 compares room temperature and cryogenic background spectra, which were taken away from the location of any emitters. At both temperatures, the SiN produces broad, featureless peaks spanning the visible range from around 550 nm to 800 nm. Both background spectra are dominated by the Si Raman line at 547.2 nm (520 cm$^{-1}$) and the 2$^{nd}$-order transverse optical (2TO) Raman peak. The Raman line at 547.2 nm occurs due to the use of a 532 nm laser for excitation. Some residual scattered laser light is also present in the background spectra. Since the Raman peaks are features of the underlying substrate, spectra taken from every location on the sample will include them.

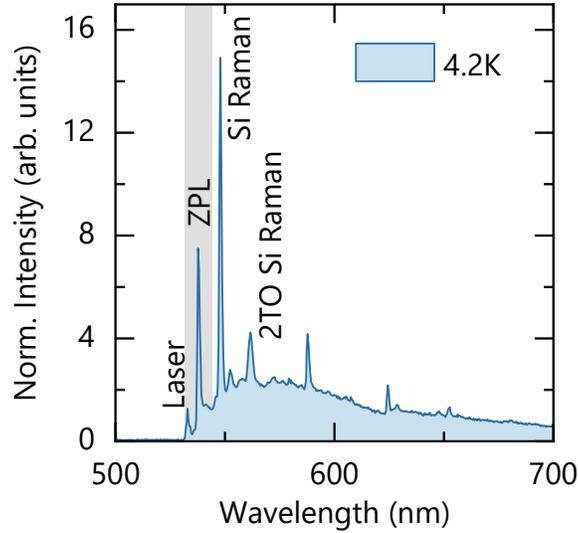

**Figure S3:** SiN SPE PL spectra without background correction.

In contrast, Figure S3 shows low temperature PL of an SPE. Both the Si Raman peaks are visible alongside the emission from the defect itself, which consists of a narrow ZPL line, broad phonon sidebands, and narrow vibronic lines discussed in the main text. Because the Si Raman and 2TO Si Raman are always present in PL spectra and the ZPLs only appear in spectra taken from the location of isolated bright spots in intensity maps, we can conclude that these are distinct features.

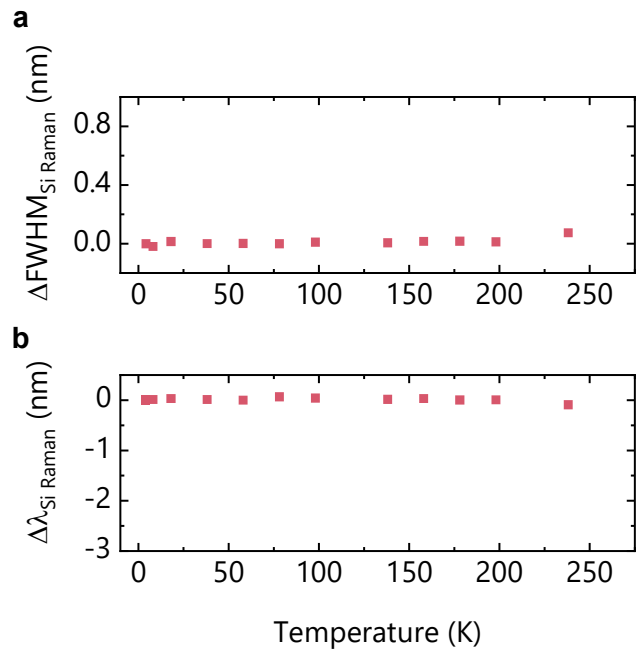

**Figure S4:** a) Change in FWHM of the Si Raman peak in background spectra vs temperature. b) Change in Si Raman peak from background spectra vs temperature.

This distinction is confirmed by Figure S4, which compares the change in FWHM (a) and peak position (b) of the main Si Raman line as a function of temperature. The vertical axes in Figure S4 have been scaled to the same magnitude of the change in SiN SPE ZPL FWHMs and peak positions, respectively, for comparison. It is clear from Figure S3 that the Si Raman peak characteristics remain effectively constant as a function of temperature. Whatever variations are present can be attributed to fitting errors or differences in the sample focus.

## 3. Additional Frames from Time-Resolved Spectra

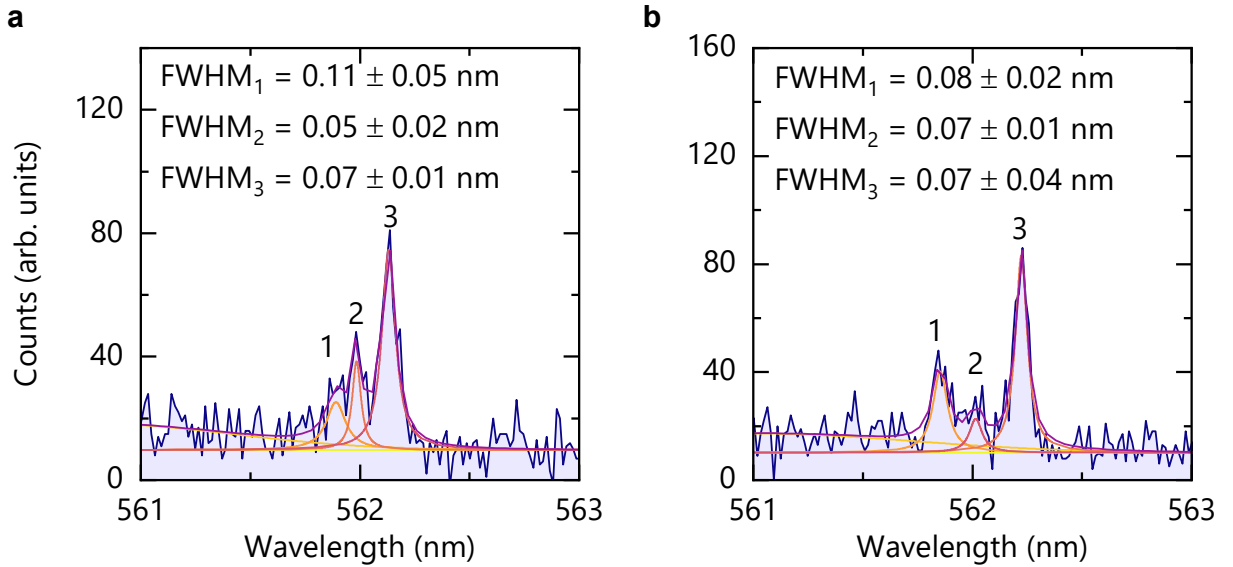

**Figure S5:** a,b) Additional frames from time-resolved spectra showing narrow, resolution-limited peaks. Linewidths from fitting each peak are shown as well.

Additional frames from high-resolution time-resolved spectra are shown in Figure S5. Again, distinct peaks from the ZPL transition energy wandering are present. These peaks can be well fitted to Lorentzian curves, and the linewidths of each peak are labeled in Figure S5. The narrowest of these ZPL peaks has a linewidth of $0.05 \pm 0.02$ nm, or $\sim 47$ GHz. Though slightly broader than the shortest peak reported in the main text, this line is still approximately spectrometer resolution-limited (0.029 nm).

## 4. Power Dependence of Spectral Diffusion

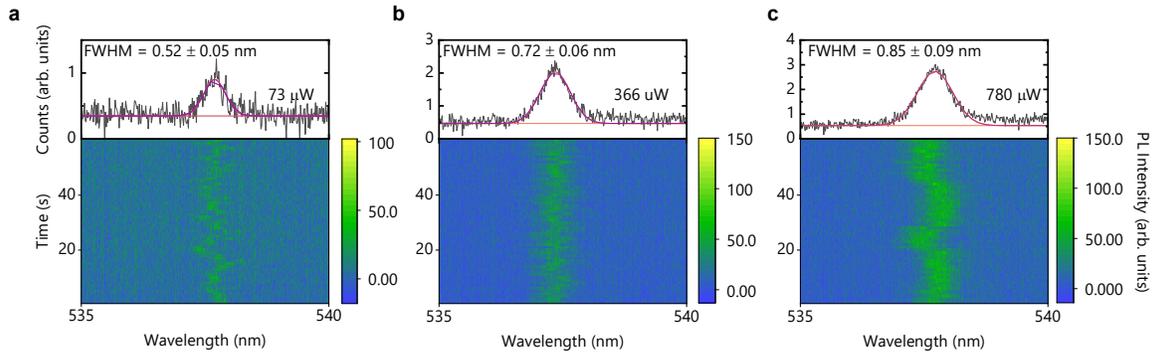

**Figure S6:** Power dependence of the ZPL linewidth. Time-resolved (bottom) and integrated (top) spectra at 73 µW (a), 366 µW (b), and 780 µW (c) excitation powers.

We also see that the ZPL inhomogeneous linewidth is power-dependent. Figure S6 a-c shows the time-resolved (bottom) and integrated (top) spectra of an emitter's ZPL at increasing excitation power. As the excitation power is increased from 73 µW to 780 µW, it is clear from the time-resolved spectra that the ZPL line broadens by greater spectral wandering. Part (c) of Figure S6, where the highest excitation power was used, exhibits abrupt jumps in the ZPL position of ~ 0.4 nm. The contribution from increased spectral diffusion is reflected the integrated spectra, where the ZPL has a linewidth of 0.52 ± 0.05 nm at 73 µW, 0.72 ± 0.06 nm at 366 µW, and 0.85 ± 0.09 nm at 780 µW (taken from Gaussian fits to the peak).

This motivates the possibility of using ultra-low excitation power to mitigate inhomogeneous broadening by spectral diffusion and obtain transform-limited emission lines at low temperatures. Excitation and outcoupling efficiencies could be significantly improved by Purcell enhancement in the form of patterning SiN films and/or coupling emitters to plasmonic nanocavities; this would allow for high enough emission at low excitation power to circumvent spectral diffusion mechanisms.

## 5. Selected SPEs

**Table S1:** ZPL peak position, inhomogeneous broadened ZPL linewidth, and $g^{(2)}(0)$ values for selected emitters in the main text. Note that time-resolved data from SPE #3 was used for Figure 3 in the main text, so the FWHM value came from fitting 4.2 K PL spectra taken with a lower-resolution grating.

| SPE No. | ZPL Wavelength (nm) | ZPL FWHM (nm) | $g^{(2)}(0)$ |
|---|---|---|---|
| 1 | 554 | 0.27 (260 GHz) | 0.41±0.04 |
| 2 | 537 | 0.49 (51 GHz) | 0.64±0.03 |
| 3 | 562 | 0.33* (31 GHz) | 0.75±0.02 |
| 4 | 545 | 0.27 (260 GHz) | 0.18±0.02 |
| 5 | 542 | 0.66 (67 GHz) | 0.41±0.07 |
| 6 | 548 | 0.56 (56 GHz) | 0.38±0.05 |
| 7 | 547 | 0.55 (55 GHz) | 0.40±0.06 |
| 8 | - | - | 0.42±0.04 |